\documentclass[12pt]{article}
\usepackage[utf8]{inputenc}
\usepackage[russian,english]{babel} 
\usepackage[T2A]{fontenc}  
\usepackage{amsfonts,amssymb,amsmath} 
\usepackage{graphicx} 
\usepackage{hyperref} 
\hypersetup{hidelinks} 
\usepackage{mathtools}
\usepackage{cite} 

\allowdisplaybreaks 
\makeatletter \@addtoreset{equation}{section}
\renewcommand\theequation
  {\ifnum \c@section>\z@ \thesection.\fi \@arabic\c@equation}
\renewcommand{\@biblabel}[1]{#1.}
\makeatother

\begin{document} 

\title{\setcounter{footnote}{0}\Large\bf Gradient Projection Method 
and Stochastic Search \\
for Some Optimal Control Models \\
with Spin Chains.~I\footnote{This research was funded by the Russian Science Foundation, project No.~22-11-00330-P, and performed at the Steklov Mathematical Institute of Russian Academy of Sciences.}}

\author{\setcounter{footnote}{6}\bf Oleg~V.~Morzhin\footnote{E-mail: \url{morzhin.oleg@yandex.ru};~ 
   \href{http://www.mathnet.ru/eng/person30382}{mathnet.ru/eng/person30382};~ 
    \href{https://orcid.org/0000-0002-9890-1303}{ORCID 0000-0002-9890-1303}} 
    \vspace{0.2cm} \\
\normalsize Department of Mathematical Methods for Quantum Technologies, \vspace{-0.1cm} \\ 
\normalsize Steklov Mathematical Institute of Russian Academy of Sciences, \vspace{-0.1cm} \\
\normalsize Gubkina Str.~8, Moscow, Russian Federation} 

\date{}
\maketitle
\vspace{-0.8cm}

\begin{abstract}
\small This article (I) considers the known optimal control model of a quantum information transfer along a spin chain with controlled external parabolic magnetic field, with an~arbitrary length. The article adds certain lower and upper pointwise constraints on controls, adds the problem of keeping the signal at the last spin, considers various classes of controls. For these problems under piecewise continuous controls, the projection-type linearized Pontryagin maximum principle, gradient projection method's constructions in its one- and two- and three-step forms were adapted by analogy with [Morzhin~O.V., Pechen~A.N. {\it J.~Phys.~A: Math. Theor.}, 2025]. Moreover, an~example with a~genetic algorithm's successful use under a~special class of controls is given.
\par В данной статье (I) рассматривается известная модель оптимального управления о передаче квантовой информации по спиновой цепочке с управляемым внешним магнитным полем, произвольной длиной. Статья добавляет конкретные нижние и верхние поточечные ограничения на управления, добавляет задачу об удержании сигнала на последнем спине, рассматривает разные классы управлений. Для этих задач при кусочно-непрерывных управлениях адаптированы проекционного типа линеаризованный принцип максимума Понтрягина, конструкции метода проекции градиента в его одно- и двух- и трехшаговых формах по аналогии с [Morzhin~O.V., Pechen~A.N. {\it J.~Phys.~A: Math. Theor.}, 2025]. Более того, дан пример с успешным применением генетического алгоритма при специальном классе управлений.
\vspace{0.15cm}\par {\bf Keywords:} quantum optimal control, gradient projection method, genetic algorithm, spin chains.
\par {\bf Ключевые слова:} квантовое оптимальное управление, метод проекции градиента, генетический алгоритм, спиновые цепочки.
\end{abstract}
\normalsize

\newpage

\rightline{\it Dedicated to the memory of Prof. V.A.~Dykhta (1949--2025)~\cite{Morzhin_AiT_2025},}
\rightline{\it whose contribution to the mathematical theory of optimal control and its} 
\rightline{\it applications is impressive, relates, in particular,  to quantum control} 

\vspace{-0.1cm} 
\section{Introduction-I}  
\vspace{-0.4cm}

~\par Quantum optimal control (QOC) is an~important scientific direction significantly using the mathematical theory of optimal control (MTOC),~etc. The large book~\cite{Tannor_book_2007} (D.J.~Tannor, 2007) is a~good introduction to quantum mechanics with a~time-dependent perspective, including for QOC. Note, e.g., the following tools used for various quantum finite- and infinite-dimensional (fin.-dim., infin.-dim.) optimization problems:

\par -- {\it conjugate gradient} and {\it Polak--Ribi{\`e}re--Polyak methods} in the large-dimensional nonconvex unconstrained optimization in $\mathbb{R}^n$ for molecular design in \cite{Yakovlev_Anikin_Bolshakova_Gasnikov_Gornov_et_al_2019} (P.A.~Yakovlev, A.S.~Anikin, O.A.~Bol'shakova, A.V.~Gasnikov, A.Yu.~Gornov, et al., 2019);

\par -- {\bf first-order GPM} for constrained maximizing an~objective function defined on a~{\it complex fin.-dim. Stiefel (sub)manifold} constructed with the {\it fin.-dim. Kraus operators} in \cite{Oza_Pechen_Dominy_et_al_JPA_2009} (A.~Oza, A.~Pechen, J.~Dominy, V.~Beltrani, K.~Moore, H.~Rabitz, 2009) where points of the Stiefel (sub)manifold are interpreted as controls (note that Kraus map is a~fundamental concept, e.g., for quantum channels, and the corresponding GPM is also a~fundamental result) (the term ``first-order'' underlines that the fin.-dim. optimization is via a~special first-order matrix ordinary differential equation (ODE));  

\par -- over various matrix sets,~{\bf one-step GPM} (GPM-1S) for minimizing the {\it Hartree--Fock functional}, etc. in \cite{Cances_Pernal_2008} (E.~Canc{\'e}s, K.~Pernal,~2008) and {\bf GPM-1S}, {\bf two-step GPM} (GPM-2S) in \cite{Bolduc_Knee_et_al_2017} (E.~Bolduc, et al., 2017) for optimization in tomography of quantum states;

\par -- {\it GRadient Ascent Pulse Engineering (GRAPE)} (\hspace{-0.05cm}\cite{Khaneja_et_al_2005} (N.~Khaneja, et al., 2005), \cite{Pechen_Tannor_2012} (A.N.~Pechen, D.J.~Tannor, 2012), etc.) (under {\it piecewise constant} (PConst.) unconstrained controls, deriving {\it fin.-dim.} gradients directly, i.e. not in the terms of any adjoint systems, etc.) (the term ``pulse'' relates to PConst. controls); 

\par -- {\it incoherent GRAPE (inGRAPE)} in \cite{Petruhanov_Pechen_JPA_2023} (V.N.~Petruhanov, A.N.~Pechen, 2023), i.e. for the concept initially developed in~\cite{Pechen_Rabitz_2006}~(A.N.~Pechen, H.~Rabitz, 2006), etc. and considering an~open quantum dynamics driven by coherent and incoherent controls; 

\par -- {\it Pontryagin maximum principle} (PMP)~(e.g., the survey~\cite{Boscain_Sigalotti_Sugny_2021}\,(U.~Boscain, M.~Sigalotti, D.~Sugny, 2021)), {\it projection-type linearized PMP} in \cite{Buldaev_Kazmin_2022} (A.S.~Buldaev, I.D.~Kazmin, 2022), \cite{Morzhin_Pechen_JPA_2025}~(O.V.~Morzhin, A.N.~Pechen, 2025); 

\par -- {\bf GPM-1S, GPM-2S, and three-step GPM} (GPM-3S) for optimizing {\it piecewise continuous (PContin.)} program controls taken in the Gorini--Kossakowski--Sudarshan--Lindblad (under the initially developed in \cite{Pechen_Rabitz_2006} concept of incoherent controls), Schr\"{o}dinger equations, with deriving the needed {\it adjoint systems} and {\it infin.-dim.} gradients, e.g., in \cite{Morzhin_Pechen_QIP_2023, Morzhin_Pechen_JPA_2025}, \cite[Subsect. 4.3]{Morzhin_Pechen_IrkutskUniv_2023} (O.V.~Morzhin, A.N.~Pechen) or {\it fin.-dim.} {\bf GPM-1S, GPM-2S} under {\it PConst. controls} in such Subsect.\,4.2 of \cite{Morzhin_Pechen_IzvMath_2023}\,(O.V.~Morzhin, A.N.~Pechen, 2023) that explicitly exactly shows how the final state, objective functions' gradients depend on those parameters that determine PConst. controls, without involving any infin.-dim. gradients, for a~two-level quantum optimal control problem (OCP) (Theorems~1,\,2 in \cite{Morzhin_Pechen_IzvMath_2023});

\par -- {\bf GPM-1S} in \cite{Pechen_LJM_2025}\,(A.N.~Pechen, 2025) under {\it PConst. controls}, with deriving a~needed fin.-dim. gradient directly, i.e. not in the terms of any adjoint systems, etc., i.e. similar to GRAPE in this meaning, and with a~specialization in the important issue to numerically analyze the landscape (i.e. graph) of a~terminal objective functional in the terms of and near a~{\it trap} of some order for the objective functional, such a~trap stops GPM-1S and is not a~globally optimal control; 

\par -- {\it Krotov-type iterative methods} (e.g., \cite[\S\,6.5]{Krotov_book_1996}\,(V.F.~Krotov, 1996), \cite[\S 16.2.2]{Tannor_book_2007}~(D.J.~Tannor, 2007), 
\cite{Caneva_Murphy_Calarco_et_al_2009}~(T.~Caneva, M.~Murphy, T.~Calarco, et al., 2009), \cite{Murphy_Montangero_Giovannetti_Calarco_2010} (M.~Murphy, S.~Montangero, V.~Giovannetti, T.~Calarco, 2010), \cite{Baturina_Morzhin_AiT_2011}\,(O.V.~Baturina, O.V.~Morzhin, 2011), \cite{Morzhin_Pechen_IrkutskUniv_2023}\,(O.V.~Morzhin, A.N.~Pechen, 2023), etc., including the {\it projection} types (under some regularization) in \cite[\S\,6.5]{Krotov_book_1996}, \cite[Subsect. 3.4]{Morzhin_Pechen_IrkutskUniv_2023}; 

\par -- {\it variational maximum principle, etc. known in the theory of impulse optimal control, i.e. with combining ``ordinary'' controls and $\delta$-pulses}, e.g., in  \cite{Dykhta_Samsonyuk_book_2nd_ed_2003} (V.A.~Dykhta, O.N.~Samsonyuk, 2003, \S\,3.7,\,6.8); 

\par -- {\it impulse iterative Krotov-type method} developed in~\cite{Krotov_Morzhin_Trushkova_2013}~(V.F.~Krotov, O.V.~Morzhin, E.A.~Trushkova, 2013) where Sect.~7 considers a~transfer problem with the Landau--Zener-type Hamiltonian; 

\par -- {\bf stochastic search} with {\bf genetic algorithm} (GA) (e.g., in \cite{Judson_Rabitz_PRL_1992}~(R.S.~Judson, H.~Rabitz, 1992), in the noted above \cite{Pechen_Rabitz_2006}), with {\it dual annealing and differential evolution} (in~\cite{Morzhin_Pechen_IzvMath_2023}, \cite{Morzhin_DYSC2025}~(Morzhin O.V.,~2025)); 

\par -- {\it Chopped RAndom-Basis (CRAB)} \cite{Caneva_Calarco_Montangero_2011} (T.~Caneva, T.~Calarco, S.~Montangero, 2011) (trigonometric terms, etc.).

As \cite{Murphy_Montangero_Giovannetti_Calarco_2010} notes, using spin chains as quantum channels for communication between two parties (sender, receiver) was first proposed in \cite{Bose_2003}\,(S.~Bose, 2003). From the point of view of the mathematical modelling for quantum information processing, an important problem is to model a successive transfer of a~quantum signal along a~spin chain (qubit arrays) (e.g., \cite{Bose_2003, Murphy_Montangero_Giovannetti_Calarco_2010, Gurman_Rasina_AiT_2014, Gurman_Guseva_Fesko_AIPConfProc_2016, Feldman_Pechen_Zenchuk_2024, Morzhin_DYSC2025} and many others). This article uses such the known from~\cite{Caneva_Murphy_Calarco_et_al_2009, Murphy_Montangero_Giovannetti_Calarco_2010} model for {\it transferring} a~single excitation along a~{\it spin-$\frac{1}{2}$ chain} that: 1)~the spin chain's length is an arbitrary~$N$, the dynamics is determined by such the $N$-level Schr\"{o}dinger equation that its  Hamiltonian represents Heisenberg-type interactions and, in addition, a~controlled external parabolic magnetic field; 2)~the goal is to concentrate the~excitation at the last spin site with representing this goal, first, in the terms of the quantum probability and, second, via the corresponding terminal objective functional (its values are not larger than~1) to be {\it maximized} under a~given final time in order to realize the goal. The known before model \cite{Balachandran_Gong_2008}\,(V.~Balachandran, J.~Gong, 2008) was in \cite{Murphy_Montangero_Giovannetti_Calarco_2010, Caneva_Murphy_Calarco_et_al_2009} extended to arbitrary (PContin.) controls with their optimizing via some adapted first-order Krotov method. For the transfer problem, \cite{Caneva_Calarco_Montangero_2011} uses the corresponding CRAB method. 

The articles \cite{Gurman_Rasina_Baturina_IFAC_2013, Gurman_Rasina_AiT_2014, Gurman_Guseva_Fesko_AIPConfProc_2016}, etc. (V.I.~Gurman, I.V.~Rasina, O.V.~Baturina, O.V.~Fesko, I.S.~Guseva, 2013--2016), \cite{Trushkova_AiT_2013}~(E.A.~Trushkova, 2013), use the squared distance in $\mathbb{C}^N$ as the objective functional and use some another Krotov-type optimization, and a~key idea in \cite{Gurman_Rasina_Baturina_IFAC_2013, Gurman_Guseva_Fesko_AIPConfProc_2016} is to combine such the optimization with the Gurman's method of derived problems which is known in MTOC. An important aspect is a~suitable {\it final time}, about which note the {\it analytical} results in \cite[Sect.\,3]{Gurman_Rasina_AiT_2014}~(V.I.~Gurman, I.V.~Rasina, 2014) where the before considered in \cite{Caneva_Murphy_Calarco_et_al_2009} transfer problems (Landau-Zener-type and the spin chain's model with $N=2$) were modified as noted above and analytically studied. A~beautifil result in \cite[Sect.\,3]{Gurman_Rasina_AiT_2014} is that a~transfer's time is not less than~$\pi/2$. After the optimization results shown in \cite{Murphy_Montangero_Giovannetti_Calarco_2010} with some shift for the location control~$u_2$, \cite{Trushkova_AiT_2013} considers such a~modification of the Hamiltonian that incorporates such a~shift. The work \cite{Morzhin_DYSC2025} (O.V.~Morzhin, 2025) and the current article also shift $u_2$ in the Hamiltonian.

In contrast to GPM-2S used in \cite{Bolduc_Knee_et_al_2017}, the articles \cite{Morzhin_Pechen_QIP_2023, Morzhin_Pechen_IrkutskUniv_2023, Morzhin_Pechen_JPA_2025} (O.V.~Morzhin, A.N.~Pechen), the talk \cite[slide 17]{Morzhin_talk_2019} (O.V. Morzhin, 2019), and the current article consider the {\it two-step gradient optimization} in the {\it dynamic} optimization problems. The mathematical fin.-dim. optimization theory --- long before \cite{Bolduc_Knee_et_al_2017} --- contains the GPM-2S, GPM-3S suggested in \cite{Antipin_1989, Antipin_DifferEqu_1994}\,(A.S.~Antipin, 1989, 1994), \cite{Nedich_1993}\,(A.~Nedich, 1993) for constrained minimization of a~smooth convex (or strongly convex) function in $\mathbb{R}^n$ with the Lipschitz continuous gradient. Moreover, note \cite[Sect.\,4]{Polyak_1969}~(B.T.~Polyak, 1969) for the {\it conjugate gradient method}'s use in 
a~constrained problem. The {\it heavy-ball method} known for unconstrained optimization~\cite{Polyak_ZVMMF_1964} (B.T.~Polyak, 1964) is an~origin of these fin.-dim. GPM forms and also of the conjugate gradient \cite{Polyak_1969}  and Polak--Ribi{\`e}re--Polyak methods used in \cite{Yakovlev_Anikin_Bolshakova_Gasnikov_Gornov_et_al_2019}. Both GPM-2S, GPM-3S are {\it not usual} for MTOC, where various GPM-1S forms are {\it long known}, e.g., from~\cite{Levitin_Polyak_1966}\,(E.S.~Levitin, B.T.~Polyak, 1966), \cite{Demyanov_Rubinov_1968_1970}\,(V.F.~Demyanov, A.M.~Rubinov, 1968), \cite{Nikolskii_2007}\,(M.S.~Nikolskii, 2007), \cite[\S\,3.3]{Buldaev_2010}\,(A.S.~Buldaev, 2010), etc. As \cite{Morzhin_Pechen_JPA_2025}, etc. numerically show for some quantum OCPs, GPM-2S, GPM-3S can be {\it significantly better} than GPM-1S. 

The structure of this article (I) is as follows. Sect.~\ref{Sect2} gives the statements of the considered OCPs. Sect.~\ref{Sect3} adapts the projection-type linearized PMP and the needed GPM constructions under PContin. controls, analytically shows the insufficiency of the zero control for transferring, keeping ($N=3$). Sect.~\ref{Sect4} gives the formulas for solving the quantum system under PConst. controls and, for the case $N=3$, shows a~good transfer with a~special introduced class of controls and with~GA.

\vspace{-0.1cm}
\section{Optimal Control Models with a Spin Chain}
\label{Sect2}
\vspace{-0.4cm}

~\par Based on \cite{Murphy_Montangero_Giovannetti_Calarco_2010, Caneva_Murphy_Calarco_et_al_2009, Trushkova_AiT_2013}, consider such a one-dimensional spin chain with $N>1$ sites that its evolution at a~given $[0,T]$ is determined as \vspace{-0.2cm}
\begin{eqnarray}
&&\hspace{-1.1cm} \frac{d\psi^u(t)}{dt} = -i \big(H_0 + H_1(t,u(t)) \big)\psi^u(t), \quad \psi^u(0) = \psi_0, ~~ u = (u_1, u_2),
\label{sect2_f1}\\
&&\hspace{-1.1cm} H_0 =  -2J \mathbb{I}_N + J \left(\begin{smallmatrix}
1 & 1 & 0 & ... & 0 & 0\\
1 & 0 & 1 & ... & 0 & 0\\
0 & 1 & 0 & ... & 0 & 0\\
...&...&...&...&...&...&\\
0 & 0 & 0 & ... & 0 & 1\\
0 & 0 & 0 & ... & 1 & 1 
\end{smallmatrix}\right) = 
\left(\begin{smallmatrix}
-1 & 1 & 0 & ... & 0 & 0\\
1 & -2 & 1 & ... & 0 & 0\\
0 & 1 & -2 & ... & 0 & 0\\
...&...&...&...&...&...&\\
0 & 0 & 0 & ... & -2 & 1\\
0 & 0 & 0 & ... & 1 & -1 
\end{smallmatrix}\right) \text{at}\,J=1, 
\label{sect2_f2}\\
&&\hspace{-1.1cm} H_1(t,u) = {\rm diag}(g_m(t,u))_{m=1}^N, ~~ g_m(t,u) = u_1 (m - 1 - \sigma(t) - u_2)^2.  
\label{sect2_f3} \vspace{-0.2cm}
\end{eqnarray}
Here $H_0$ and $H_1$ are correspondingly the time-independent free Hamiltonian and time-dependent controlling Hamiltonian (as constant real sparse matrices), the interspin coupling $J$ in the Heisenberg-type $H_0$ and the Planck constant are already taken to be~1. We have the vector ODE with complex-valued state $\psi^u(t)$ with the invariant $\| \psi^u(t)\|_{\mathbb{C}^N}=1$, i.e. $\psi^u(t) \in S^{2N-1}(1) \subset \mathbb{C}^N$. As it is described in \cite{Murphy_Montangero_Giovannetti_Calarco_2010}, the dimension of the effective Hilbert space is simply $N$, i.e. not $2^N$. Program real-valued controls $u_1$ and $u_2 = d - \sigma$ describe correspondingly the {\it intensity} and shifted {\it position} ($d(t)$ controls a~position in~\cite{Murphy_Montangero_Giovannetti_Calarco_2010}) of an external magnetic parabolic field and are considered below (for the developed concept) in the following various forms and taking into account a~given~$\sigma$.  
\begin{itemize}
\item Let control $u$ as {\it PContin.}, the shift $\sigma(t) = w\,t$ with $w=(N-1)/T$, and, in addition to the known model, the pointwise constraint
\vspace{-0.15cm}
\begin{equation}
u(t)=(u_1(t), u_2(t)) \in Q_u(t) := [-b_1(t), \, b_1(t)] \times [-b_2(t), \, b_2(t)] \vspace{-0.2cm}  
\label{sect2_f4} 
\end{equation} 
is introduced at $t \in [0,T]$, where, by analogy with \cite[Eq.\,(4)]{Morzhin_Pechen_JPA_2025} considering some another OCPs, take the concave functions $b_l = \overline{b}_l\,{\rm sinc}\,\big(2^{q_l} \pi (t/T - 1/2)^{q_l} \big)$, $l=1,2$, where $\overline{b}_l>0$ and, e.g., $q_l=8$, so that $b_l(0)=b_l(T)=0$, $l=1,2$ for the considered here requirement of {\it smooth switching on and off} of $u_1, u_2$ with $u_l(0)=u_l(T)=0$, $l=1,2$.

\vspace{-0.1cm}

\item Let control $u$ as {\it PConst.}, the shift $\sigma(t) = w \sum\nolimits_{j=0}^{M-1} \theta_{[t_j, t_{j+1})}(t) (t_j+t_{j+1})/2$, where $w=(N-1)/T$, $\theta_{[t_j, t_{j+1})}$ is the characteristic function, a~given time grid $\Theta_M = \{t_0 = 0$, $t_1 = (\Delta t)_1$, $t_2 = t_1 + (\Delta t)_2$, ..., $t_{M-1} = T - (\Delta t)_M$, $t_M = T\}$ (as a variant, the grid can be uniform with $\Delta t=T/M$), vector ${\bf a} = (a_1, ..., a_s, ..., a_{2M}) = \big(c_{1,1}$, $...,$ $c_{1,M},$ $c_{2,1},$ $...,$ $c_{2,M}\big)$ with $c_j = (c_{l,j})_{l=1}^{2} = u(t_{j-1})$ ($j \in \overline{1,M}$) (i.e. the values of~$u$ along all the partial ranges at $[0,T]$), and the constraint \vspace{-0.2cm}
\begin{equation}
{\bf a} \in Q_{\bf a} = (\bigtimes\nolimits_{j=1}^M [-\nu_{1,j}, \nu_{1,j}]) \bigtimes (\bigtimes\nolimits_{j=1}^M [-\nu_{2,j}, \nu_{2,j}]) \subset \mathbb{R}^{2M}. 
\label{sect2_f5} \vspace{-0.2cm}
\end{equation} 
Here we define $\nu_{l,j}$ from $b_l$. If we want that the ``shelves'' $\{\nu_{l,j}\}$ should (almost) not violate the original constraint~(\ref{sect2_f4}), then we take into account that the functions $b_l$, $l=1,2$ first quickly tend from 0 to\,1, then change very slowly, and then quickly decrease to~0. However, if $\Delta t=T/M$ is sufficiently small, then one can simply define $\nu_{l,j}=b_l(t_j + \Delta t/2)$. Let $r \in \overline{1,M}$ to show which particular range is taken. For some $s$th element of~${\bf a}$, we have \vspace{-0.2cm}
\begin{equation*}
r = \begin{cases}
s~{\rm mod}~M, & s~{\rm mod}~M \neq 0, \\
M, & s~{\rm mod}~M = 0,
\end{cases}~~
l = \begin{cases}
\lfloor s/M \rfloor +1, & s~{\rm mod}~\neq 0, \\
\lfloor s/M \rfloor, & s~{\rm mod}~M = 0.
\end{cases}\vspace{-0.2cm}
\end{equation*} 

\vspace{-0.1cm}

\item Consider $\sigma = w\,t$ and {\it piecewise linear} (PLinear) control~$u$ at a~given grid $\Theta_M$. This case of controls' interpolation is in order to a~possible realization of the infin.-dim. gradients given below in Lemma~1. 

\item The following {\it special class of continuous} controls is introduced and represents some possible non-instantaneous pulses near $t=0,T$: \vspace{-0.2cm}
\begin{equation}
Q_u(t) \ni u_l(t) = \begin{cases}
\theta^{\rm L}_l b_l(t), & t \in [0, \widehat{t}_1), \\
\frac{y_l - \theta^{\rm L}_l b_l(\widehat{t}_1)}{(\widehat{t}_2- \widehat{t}_1)^2}(t - \widehat{t}_1)^2 + \theta^{\rm L}_l b_l(\widehat{t}_1), & t \in [\widehat{t}_1, \widehat{t}_2), \\
y_l, & t \in [\widehat{t}_2, \widehat{t}_3), \\
\frac{y_l - \theta^{\rm R}_l b_l(\widehat{t}_4)}{(\widehat{t}_4 - \widehat{t}_3)^2}(t - \widehat{t}_4)^2 + \theta^{\rm R}_l b_l(\widehat{t}_4), & t \in [\widehat{t}_3, \widehat{t}_4), \\
\theta^{\rm R}_l b_l(t), & t \in [\widehat{t}_4, T],
\end{cases}  \vspace{-0.2cm}  
\label{sect2_f6}
\end{equation}
where $l=1,2$, $0 < \widehat{t}_1 < \widehat{t}_2 < \widehat{t}_3 < \widehat{t}_4 < T$, ``L'' and ``R'' mean ``left'' and ``right''. In this class, consider ${\bf x} = \big(\theta_1^{\rm L}, \theta^{\rm R}_1$, $\theta_2^{\rm L}, \theta^{\rm R}_2$, $y_1, y_2, \widehat{t}_1, \widehat{t}_2, \widehat{t}_3, \widehat{t}_4\big)$ as controlling parameters' vector in a~given compact $Q_{\bf x}$ formed, e.g., with $\theta_l^{\rm L}, \theta_l^{\rm R} \in [-1,1]$, $y_l \in [-0.1 \overline{b}_l, 0.1 \overline{b}_l]$, $l=1,2$, and $\widehat{t}_1 \in [\xi_1, \xi_2]$, $\widehat{t}_2 \in [\xi_3, \xi_4]$, $\widehat{t}_3 \in [\xi_5, \xi_6]$, $\widehat{t}_4 \in [\xi_7, \xi_8]$ with such given $\{\xi_s\}_{s=1}^8$ that  $0<\xi_1<\xi_2<\xi_3<\xi_4 \leq 0.2T$, $0.8T \leq \xi_5<\xi_6<\xi_7<\xi_8<T$. This class is inspired by the shown in Fig.~(a),(b) in \cite{Morzhin_DYSC2025}\,(O.V.~Morzhin, 2025) some another special class of controls proposed by the author. In order to solve (\ref{sect2_f1})--(\ref{sect2_f3}) with $u$ determined by (\ref{sect2_f6}), we, as a~variant, approximate this $u$ via its {\it PConst.} version at a~given grid $\Theta_M$, where we take $u_l(t_j)$ at each $[t_j, t_{j+1}]$. 

\end{itemize}  
In the PConst. case of $u$ and $\sigma$, the dynamic system is discrete-continuous. Here the given below Lemma~2 gives the exact solution of the quantum ODE simply as the chronological product with the unitary propagators and $\psi_0$. E.g., the theory of Magnus expansion from the theory of ODEs is not needed here, because the coefficients of the considered ODE are constant. There is an~analogy with the corresponding stage of GRAPE, etc. In contrast, \cite{Morzhin_DYSC2025} uses the problem's realification and the Radau method. 

Consider the problem of {\it transfer} $\psi^u(0) = \psi_0 = (1,0, \dots, 0)^{\rm T}$ to $\psi_{\rm g}=(0, \dots, 0,1)^{\rm T}$ in the meaning to make the final probability~$\big|\langle \psi^u(T), \psi_{\rm g}\rangle \big|^2_{\mathbb{C}^N}$ equal to~1 (with a~high precision). A~possible way is, as in \cite{Murphy_Montangero_Giovannetti_Calarco_2010, Caneva_Murphy_Calarco_et_al_2009, Trushkova_AiT_2013}, to consider an~assigned $T$ and the corresponding problem to maximize the final probability over a~class $\mathcal{U}$ of admissible controls, i.e. \vspace{-0.2cm}
\begin{eqnarray}
\hspace{-0.7cm}I_1(u) &=& F(\psi^u(T);\psi_{\rm g})= 1 - \big|\langle \psi^u(T), \psi_{\rm g}\rangle \big|^2 =\nonumber\\
&=&1 - |\psi_N^u(T)|^2 = 1 - ({\rm Re}\, \psi_N(T))^2 - ({\rm Im}\, \psi_N(T))^2 \to \min\limits_{u \in \mathcal{U}},
\vspace{-0.25cm} 
\label{sect2_f7} 
\end{eqnarray}
and, if it is needed, to change~$T$. The geometry of $I_1(u)$ is determined by the function $F(\psi;\psi_{\rm g})$, the dynamic equation, etc. In the view of $\|\psi\|=1$, one has that $F(\psi;\psi_{\rm g}) \in [0,1]$. The minimal value of $F(\psi;\psi_{\rm g})$, i.e.~0, is reached at $\psi = (0, ..., 0, \psi_N)^{\rm T}$ with $\psi_N$ being at the complex unit circle~$S^1(1)$. If one takes $\psi = \widehat{\psi} e^{i\varphi}$ with $\varphi \in [0,2\pi)$, then $|\psi_N|^2 = |\widehat{\psi}_N|^2 |e^{i\varphi}|^2 = |\widehat{\psi}_N|^2$, i.e. $F$ is invariant with respect to the phase factor. At $t=0$, one has the largest value $F(\psi_0;\psi_{\rm g}) = 1 - |\langle \psi_0, \psi_{\rm g}\rangle|^2 = 1$. 

Of course, one can expect such a~case that, under some given~$T$, etc., there is such a~control process $(\widehat{u},\psi^{\widehat{u}})$ that is {\it globally optimal} in this OCP from the point of view of MTOC, --- i.e. the objective functional $I_1$ cannot be decreased more via any way, --- but $F(\psi^{\widehat{u}}(T); \psi_{\rm g})$ is {\it significantly larger} than~0. One can try to search such a~fixed $T$ that is the smallest one at some possible finite set of such $T$ values that $I_1 \approx 0$ for each of them. One can formulate the corresponding time-optimal control problem with changeable $T$ considered in $\Upsilon(u,T) = T \to \min$ and $\big|\langle \psi^u(T), \psi_{\rm g}\rangle |^2 = 1$. Below GPM is considered for the problem to minimize $I_1(u)$ with a~given~$T$. 

By analogy with the keeping problem given in \cite[Subsect. 4.8]{Morzhin_Pechen_JPA_2025} for some another quantum dynamics, etc., introduce --- for the spin chain's system ---  such a~given $\psi_0$ that $|\langle \psi_0, \psi_{\rm g} \rangle|^2 = 1$ and the problem to {\it keep} $\psi_0$ in the meaning of the invariant $|\langle \psi^u(t), \psi_0 \rangle|^2 \equiv 1$ at a~given $[0,T]$. At $t=0$, we have $F(\psi_{\rm g};\psi_{\rm g})=0$. As a~variant, let us to take $\psi_0 = \psi_{\rm g}$~and \vspace{-0.2cm}
\begin{equation}
\hspace{-0.0cm}\qquad I_2(u) = F(\psi^u(T); \psi_{\rm g}) + P_{\psi}\, \int\nolimits_0^T 
F(\psi^u(t); \psi_{\rm g}) dt \to \min\limits_{u \in \mathcal{U}}, ~~ P_{\psi}>0. 
\hspace{-0.0cm}
\vspace{-0.2cm}
\label{sect2_f8}
\end{equation}

Under PContin. $u$ and in addition to the time-dependent constraint (\ref{sect2_f4}), note the taken below integral term for $I_p(u)$, $p \in \{1,2\}$:
\vspace{-0.2cm}
\begin{equation}
\Phi_p(u) = I_p(u) + \int\nolimits_0^T \sum\nolimits_{l=1}^2 P_{u_l} S_l(t) u_l^2(t) dt \to \min\limits_{u \in \mathcal{U}},
\label{sect2_f9} \vspace{-0.2cm}
\end{equation}
where $P_{u_l} \geq 0$, $l=1,2$, the functions  $S_l(t) = \exp[C_{S_l}(t/T - 1/2)^2]$ with a~sufficiently large $C_{S_l}>0$,~$l=1,2$ (with respect to (\ref{sect2_f4}), it is for an~additional adjusting $u_1, u_2$ near $t=0,T$ and is by analogy with \cite[Eq.~(5)]{Morzhin_Pechen_JPA_2025} considering some another models). For the fin.-dim. GPM developed in Article~II continuing this article, consider PConst. controls in the meaning of $f_p({\bf a})= \Phi_p(u)$ to be minimized in view of (\ref{sect2_f9}) and under (\ref{sect2_f5}), $p \in \{1,2\}$, where assign a~certain {\it PConst.} $S_l$, $l=1,2$ at some grid~$\Theta_M$, as well as~$\sigma$. For a~certain PConst. approximation of~(\ref{sect2_f6}), operate with ${\bf a}$ in order to solve (\ref{sect2_f1}) via the given below in Lemma~2 way and optimize ${\bf x}$ as for \vspace{-0.2cm}
\begin{equation}
f_3({\bf x}) = I_1(u(\cdot;{\bf x})) + P_{\bf x} \sum\nolimits_{j=0}^{M-1}\sum\nolimits_{l=1}^2 |u_l(t_j;{\bf x})| \to \min\limits_{{\bf x} \in Q_{\bf x}},~~ P_{\bf x} \geq 0. \vspace{-0.2cm} 
\label{sect2_f10}
\end{equation} 
Further, for Article~II and the keeping requirement at $[0,T]$, take \vspace{-0.2cm}
\begin{equation}
u(t)=\Big({\rm Pr}_{Q_u(t)}\Big(\sum\nolimits_{i=1}^{M_{\sin}} \gamma_{l,i} \sin\big(\lceil \omega_{l,i} \rceil \pi t/T \big) \Big), ~  l=1,2 \Big), ~~ t \in [0,T] 
\label{sect2_f11} 
\vspace{-0.2cm}
\end{equation}
with the orthogonal projection, ${\bf y}=(\gamma_{1,1}, \omega_{1,1}, ..., \gamma_{2,M_{\sin}}, \omega_{2,M_{\sin}}) \in Q_{\bf y}$ (compact set), approximate $u$ at $\Theta_M$ as PConst., introduce $P_{\bf y}\geq 0$ and \vspace{-0.15cm}
\begin{equation}
f_4({\bf y}) = \max\limits_{j \in \overline{1,M}} F(\psi^{u}(t_j); \psi_{\rm g}) + 
P_{\bf y} \sum\nolimits_{j=0}^{M-1} \sum\nolimits_{l=1}^2 |u_l(t_j; {\bf y})| \to \min\limits_{y \in Q_{\bf y}}.
\label{sect2_f12}
\end{equation}

\vspace{-0.1cm}
\section{Infinite-Dimensional Gradients, Projection-Type \\Linearized PMP, Zero Control, and GPM Forms}
\label{Sect3}
\vspace{-0.4cm}

~\par{\bf Lemma 1.}
\label{lemma1}
{\it For the system (\ref{sect2_f1})--(\ref{sect2_f3}) with a~given chain's length~$N$, PContin. control $u=(u_1,u_2)$, a~given final time $T$, under the constraint~(\ref{sect2_f4}), consider the transfer problem, i.e. with $\Phi_1(u)$, or the  keeping problem, i.e. with $\Phi_2(u)$. For these problems, the following constructions at an arbitrary admissible control~$u^{(k)}$ are hold (the index ``p'' for the quantum and adjoint systems' solutions is omitted): 1)~the adjoint equation \vspace{-0.2cm}
\begin{eqnarray}
\hspace{-0.7cm}
&&\frac{d \eta^{u^{(k)}}(t)}{dt} = -i \left( H_0 + H_1(t,u^{(k)}(t)) \right) \eta^{u^{(k)}}(t) +\nonumber\\
\hspace{-0.7cm}  &&\quad + 
\begin{cases}
0, & \hspace{-0.1cm} \Phi=\Phi_1,\\
P_{\psi}\langle \psi_{\rm g},\psi^{u^{(k)}}(t) \rangle \psi_{\rm g} = (0, \dots, 0, P_{\psi}  (\psi_N^{u^{(k)}}(t))^{\ast})^{\rm T}, & \hspace{-0.1cm} \Phi=\Phi_2;   
\end{cases} 
\label{sect3_f1} 
\end{eqnarray} 
2)~the final-time condition \vspace{-0.2cm}
\begin{equation}
\eta^{u^{(k)}}(T) = \eta^{u^{(k)}}_{\rm Transv.} = \langle \psi_{\rm g}, \psi^{u^{(k)}}(T) \rangle \psi_{\rm g} = (0, \dots, 0, (\psi_N^{u^{(k)}}(T))^{\ast})^{\rm T};
\label{sect3_f2} \vspace{-0.1cm}
\end{equation}
3)~the infin.-dim. gradients ($p \in \{1,2\}$) \vspace{-0.2cm}
\begin{eqnarray}
&&\hspace{-0.2cm}{\rm grad}\, \Phi_p(u^{(k)})(t) = \Big(-{\rm Im}\Big\langle \eta^{u^{(k)}}(t),\, \frac{\partial H_1(t,u)}{\partial u_l}\Big|_{u=u^{(k)}(t)} \psi^{u^{(k)}}(t) \Big\rangle_{\mathbb{C}^N} + \nonumber\\
&&\quad\quad\quad\quad\quad  + 2 P_{u_l} S_l(t) u^{(k)}_l(t), ~ l = 1,2 \Big),~~ t \in [0,T], \vspace{-0.1cm} 
\label{sect3_f3} 
\end{eqnarray}
where $\frac{\partial H_1}{\partial u_1} = {\rm diag}\{(m-1-\sigma(t)-u_2)^2\}_{m=1}^N$, $\frac{\partial H_1}{\partial u_2} = -2\, {\rm diag}\{ u_1 (m-1-\sigma(t)-u_2)\}_{m=1}^N$; 4)~the gradient increment formula with~some~residual, i.e. \vspace{-0.1cm}
\begin{equation}
\hspace{-0.0cm}\Phi_p(u) - \Phi_p(u^{(k)}) = \langle {\rm grad}\, \Phi_p(u^{(k)}), u - u^{(k)} \rangle_{L_2([0,T], \mathbb{R}^2)} + r, ~ p \in \{1,2\}.\hspace{-0.21cm}
\label{sect3_f4} \vspace{-0.1cm}
\end{equation} }

{\bf Proof.} By analogy with \cite{Morzhin_Pechen_JPA_2025}, these constructions are derived with adapting the Krotov Lagrangian known, e.g., from \cite{Krotov_book_1996} and taken here at an admissible $(u,\psi^u)$ and with a~solving linear function $\varphi(t,\psi)={\rm Re}\langle \eta(t), \psi \rangle$ (this $\varphi$ is taken as in \cite[p. 256]{Krotov_book_1996}):
\begin{align*}
KL_p(u) = \Phi_p(u)= G(\psi^u(T)) - \int\nolimits_0^T R(t,\eta^u(t), \psi^u(t), u(t))dt
\end{align*} 
with 
\begin{align*}
G(\psi^u(T)) &= F(\psi^u(T)) + {\rm Re}\langle \eta^u(T), \psi^u(T)\rangle - {\rm Re}\langle \eta^u(0), \psi^u(0)\rangle,\\
R(t,\eta^u(t), \psi^u(t), u(t)) &= {\rm Re}\left[\langle \eta^u(t), -i(H_0 + H_1(t,u(t))) \psi^u(t)\rangle + \langle \frac{d\eta^u(t)}{dt}, \psi^u(t) \rangle \right] - \nonumber \\ 
&\quad - P_{\psi} F(\psi^u(t);\psi_{\rm g}) - \sum\nolimits_{l=1}^2 P_{u_l} S_l(t) u_l^2(t)
\end{align*} 
with $\eta^u(t)$ to be defined, and $P_{\psi}=0$, if $p = 1$. The Pontryagin function is $h(t,\eta,\psi,u)={\rm Re}\langle \eta, -i(H_0 + H_1(t,u)) \rangle - P_{\psi} F(\psi; \psi_{\rm g}) - \sum\nolimits_{l=1}^2  P_{u_l} S_l(t) u_l^2$, where $P_{\psi}=0$ beyond the keeping problem. At an arbitrary admissible $u$ and a~given admissible $u^{(k)}$, consider $KL_p(u) - KL_p(u^{(k)}) = G(\psi^u(T)) - G(\psi^{u^{(k)}}(T)) - \int\nolimits_0^T [R(t, \eta^{u^{(k)}}(t), \psi^u(t), u(t)) - R(t, \eta^{u^{(k)}}(t), \psi^{u^{(k)}}(t), u^{(k)}(t))]dt$, where $\psi^{u^{(k)}}$ is the solution of~(\ref{sect2_f1}) with $u=u^{(k)}$, and $\eta^{u^{(k)}}$ is determined below. By expanding the increment for $G$ into the matrix Taylor series up to the 1st order and equating the 1st term to zero, (\ref{sect3_f2}) is derived. Further, (\ref{sect3_f1}) is derived by expanding the increment for $R$ and equating to zero such the part of the 1st term that relates to $\psi^u(t) - \psi^{u^{(k)}}(t)$. The rest of the increment provides~(\ref{sect3_f4}),\,(\ref{sect3_f3}).~$\square$. 

This lemma is by analogy with~\cite[Sect.\,2]{Morzhin_Pechen_JPA_2025} and takes into account that \cite{Murphy_Montangero_Giovannetti_Calarco_2010} already gives the transversality condition for~(\ref{sect2_f7}). 

\vspace{0.1cm}

{\bf Corollary 1.}
\label{corollary1}
{\it In the transfer problem, consider the control $u = 0$. Then the solution of (\ref{sect2_f1}) is $\psi^{u=0}(t) = e^{-i H_0 t}\psi_0$, the solution of (\ref{sect3_f1}),~(\ref{sect3_f2}) is $\eta^{u=0}(t) = e^{i H_0 (T - t)} \eta^{u=0}_{\rm Transv.}$, and the gradient (\ref{sect3_f3}) becomes \vspace{-0.2cm}
\begin{align*}
{\rm grad}\, \Phi_p(0)(t) = \Big(\hspace{-0.15cm}-{\rm Im}\,\big\langle e^{i H_0 (T-t)} \eta^{u=0}_{\rm Transv.}, \frac{\partial H_1(t,u)}{\partial u_l} \Big|_{u=0} e^{-i H_0 t}\psi_0 \big\rangle, l=1,2 \Big), 
\end{align*}
where $\frac{\partial H_1(t,u)}{\partial u_1} \Big|_{u=0} = {\rm diag}\{(m-1 - \sigma(t))^2 \}_{m=1}^N$ and $\frac{\partial H_1(t,u)}{\partial u_2} \Big|_{u=0}=0_N$. }

{\bf Proof.} The gradient's formula at $u = 0$ is obtained by substitution of $\psi^{u=0}(t)$, $\eta^{u=0}(t)$, and $u=0$ in (\ref{sect3_f3}). For obtaining $\eta^{u=0}(t) = e^{i H_0 (T - t)} \eta^{u=0}_{\rm Transv.}$, one has $e^{-i H_0 T} c = \eta^{u=0}_{\rm Transv.}$, multiples it by $e^{i H_0 T}$, obtains $e^{i H_0 T} e^{-i H_0 T} c = \mathbb{I}_N c$ using that the matrices $i H_0 T$, $-i H_0 T$ commute and matrix exponentials are invertible matrices.~$\square$ 

\vspace{0.1cm}

By analogy with the projection-type linearized PMP known in MTOC and used for QOC, e.g., in~\cite{Buldaev_Kazmin_2022, Morzhin_Pechen_JPA_2025}, the following theorem is considered. 

\vspace{0.15cm}

{\bf Theorem 1.} (Projection-type linearized PMP).  
{\it Consider the transfer and keeping problems (\ref{sect2_f1})--(\ref{sect2_f4}), (\ref{sect2_f6})--(\ref{sect2_f8}) with PContin. $u$. For any OCP among these problems, if an admissible control $\widetilde{u}$ is a~local minimum point for the corresponding $\Phi_p(u)$, $p \in\{1,2\}$, then for $\widetilde{u}$ there exist such the solutions $\psi^{\widetilde{u}}, \eta^{\widetilde{u}}$ of (\ref{sect2_f1}),  (\ref{sect3_f1}),\,(\ref{sect3_f2}) that there is the pointwise condition \vspace{-0.2cm}
\begin{equation}
\widetilde{u}(t) = {\rm Pr}_{Q_u(t)}\left(\widetilde{u}(t) - \alpha \, {\rm grad}\, \Phi_p(\widetilde{u})(t) \right) \quad \forall~ \alpha>0, ~t \in [0,T].  
\label{sect3_f5}
\end{equation}  } 

\vspace{0.15cm}

An extremal control $\widetilde{u}$ is determined by~(\ref{sect3_f5}) for any $\alpha>0$. 

\vspace{0.15cm}

\begin{figure}[h!]
\centering
\includegraphics[width=0.5\linewidth]{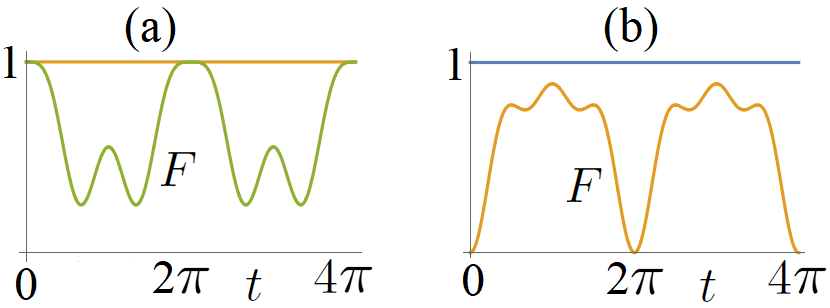}
\caption{$F(\psi^{u=0}(t); \psi_{\rm g})$ vs $t$: (a)~in the transfer problem (Example~1, $\psi_0 \neq \psi_{\rm g}$); 
(b)~in the keeping problem (Example~2, $\psi_0 = \psi_{\rm g}$).}
\label{Fig1} 
\end{figure}

{\bf Example 1.} 
Consider the transfer problem with $\Phi_1(u) = I_1(u)$ and the length $N=3$. Here,  following (\ref{sect2_f2}),~(\ref{sect2_f3}), take $H_0 =  
\left(\begin{smallmatrix}
-1 & 1 & 0\\
1 & -2 & 1\\
0 & 1 & -1
\end{smallmatrix}\right)$, $H_1 = u_1\,{\rm diag}\big((\sigma(t)+u_2)^2, (1-\sigma(t)-u_2)^2, (2-\sigma(t)-u_2)^2 \big)$. Following Corollary~\ref{corollary1}, one analytically obtains ${\rm grad}\,\Phi_1(0)(t) = \big(\text{non-zero large  expression}, 0 \big)$, i.e. $u\equiv 0$ at the whole $[0,T]$ for any~$T>0$. The second component of the vector condition (\ref{sect3_f5}) is $0=0$ at the whole $[0,T]$. Substituting the solution $\psi^{u=0}(t) = e^{-i H_0 t} \psi_0$ into the function $F(\psi; \psi_{\rm g})$ instead of $\psi$ gives $F_1 = \frac{1}{18}\big(6 \cos t + 3\cos(2t) - 2\cos(3t)+11 \big)$ which is shown in Fig.~\ref{Fig1}(a) for $t \in [0, 4\pi]$. Here we see that $u = 0$ {\it cannot} provide the transfer.~$\square$ 

\vspace{0.15cm}

{\bf Example 2.}  
For the keeping problem, the quantum system with the condition $\psi^{u=0}(0)=\psi_{\rm g}$ has the solution $\psi^{u=0}(t) = e^{-i H_0 t}\psi_{\rm g}$ whose substitution into $F(\psi; \psi_{\rm g})$ gives $F = 1-|\langle \psi^{u=0}(t); \psi_{\rm g}\rangle|^2 = \frac{2}{9}(7\cos t + 2\cos(2t) + 9)\sin^2\frac{t}{2}$, which reaches 0 at $t = 2\pi n$, $n=1,2, ...$, and is shown in Fig.~\ref{Fig1}(b) for $t \in [0, 4\pi]$. Here we see that $u = 0$ {\it cannot} provide the keeping.~$\square$ 

\vspace{0.15cm}

In this article (I), let us briefly consider the following iterative GPM-$j$S forms, $j=1,2,3$, with the infin.-dim. gradients: \vspace{-0.2cm}
\begin{align}
&\text{GPM-1S:~~\,} u_l^{(k+1)}(t) = {\rm Pr}_{[-b_l(t), \,b_l(t)]}\big(u_l^{(k)}(t) - \alpha_k \, {\rm grad}\, \Phi_p(u^{(k)})(t) \big), 
\label{sect3_f6} \\
&\text{GPM-2S:~~\,} u_l^{(k+1)}(t) = {\rm Pr}_{[-b_l(t), \,b_l(t)]}\big(u_l^{(k)}(t) - \alpha_k \, {\rm grad}\, \Phi_p(u^{(k)})(t) + \nonumber \\
&\qquad\qquad\qquad\qquad\qquad + \beta_k (u_l^{(k)}(t) - u_l^{(k-1)}(t)) \big), ~~k \geq 1, 
\label{sect3_f7} \\
&\text{GPM-3S:~~\,} u_l^{(k+1)}(t) = {\rm Pr}_{[-b_l(t), \,b_l(t)]}\big(u_l^{(k)}(t) - \alpha_k \, {\rm grad}\, \Phi_p(u^{(k)})(t) + \nonumber \\
&\qquad + \beta_k (u_l^{(k)}(t) - u_l^{(k-1)}(t)) + \gamma_k (u_l^{(k-1)}(t) - u_l^{(k-2)}(t))\big), ~~k \geq 2,  \label{sect3_f8} 
\end{align}
where $l \in \{1,2\}$, $t \in [0,T]$, GPM-2S starts from $u^{(1)}$ obtained with GPM-1S whose start is from~$u^{(0)}$, etc., the parameters $\alpha_k >0$, $\beta_k, \gamma_k \in [0,1)$ are experimentally adjusted by some ways. For (\ref{sect3_f6})--(\ref{sect3_f8}),  one can take PLinear interpolation for controls, linear $\sigma$, and PLinear representation of $b_l, S_l$, $l \in \{1,2\}$, and solve (\ref{sect2_f1}), e.g., via its realification and some numerical method, e.g., {\it the 5(4) Runge--Kutta method} with an~additional time~grid. 

\vspace{-0.1cm}
\section{System's States Represented via Matrix \\Exponentials. Stochastic Search for \\the Transfer Problem with $N=3$}
\label{Sect4}
\vspace{-0.4cm}  

~\par{\bf Lemma 2.} 
{\it Consider the system (\ref{sect2_f1})--(\ref{sect2_f3}) with a~given grid~$\Theta_M$ and, correspondingly, PConst.~$u$ and the certain PConst.~$\sigma$. At a~given admissible PConst. $\widetilde{u}$ and the corresponding~$\widetilde{\bf a}$, we have \vspace{-0.2cm}
\begin{eqnarray}
&&\hspace{-0.7cm} \psi^{\widetilde{u}}(t_j) = e^{A(\widetilde{c}_j)(t_j - t_{j-1})} \psi^{\widetilde{u}}(t_{j-1}), ~~ j \in \overline{1,M},~~ \psi^{\widetilde{u}}(t_0) = \psi_0, 
\label{sect4_f2}\\
&&\hspace{-0.7cm} A(\widetilde{c}_j) = -i\Big(H_0 + \widetilde{c}_{1,j} \,{\rm diag}\big(\big(m-1 - \frac{t_j+t_{j+1}}{2}w - \widetilde{c}_{2,j}\big)^2\big)_{m=1}^N \Big) \vspace{-0.2cm}  \label{sect4_f3}  
\end{eqnarray}
and $\psi^{\widetilde{u}}(t_j)$ ($j \in  \overline{1,M}$) depends continuously on the corresponding $\{\widetilde{{\bf a}}_s\}$.}

{\bf Proof.} The formulas (\ref{sect4_f2}), (\ref{sect4_f3}) represent the construction of the continuous function $\psi^{\widetilde{u}}$ at $[0,T]$ via such the deriving the sequential parts of this function along all the particular time ranges that each next part continues the previous part without a jump. The sequential Cauchy problems $\frac{d \psi^{\widetilde{c}_{j+1}}(t)}{dt} = A(\widetilde{c}_{j+1}) \psi^{\widetilde{c}_{j+1}}(t)$, $\psi^{\widetilde{c}_{j+1}}(t_j) = \begin{cases}
\psi^{\widetilde{c}_j}(t_j), & j \neq 0,\\
\psi_0, & j = 0,
\end{cases}$
where $\psi^{\widetilde{c}_j}(t_j)$ is taken to connect the neighboring parts of the composite trajectory, $j = \overline{0,M-1}$, are solved. The continuity of $\psi^{\widetilde{u}}(t_j)$ in the corresponding $\{\widetilde{{\bf a}}_s\}$ is based on considering the matrix exponentials as converging series.~$\square$

\vspace{0.2cm}

This lemma is needed for Example 3 in order to minimize $g_1({\bf x}) = I_1(u)$ via GA under a~PConst. approximation of~(\ref{sect2_f6}). The given by Lemma~2 procedure reminds \cite[p.\,44]{Tyatyushkin_book_1992}\,(A.I.~Tyatyushkin, 1992) (real states, beyond QOC) and the corresponding stage of GRAPE. GA is a stochastic zeroth-order optimization tool, its key ideas are shown, e.g., in \cite{Judson_Rabitz_PRL_1992, Solgi_Genetic_algorithm_Python}. 

\begin{figure}[h!]
\centering
\includegraphics[width=0.95\linewidth]{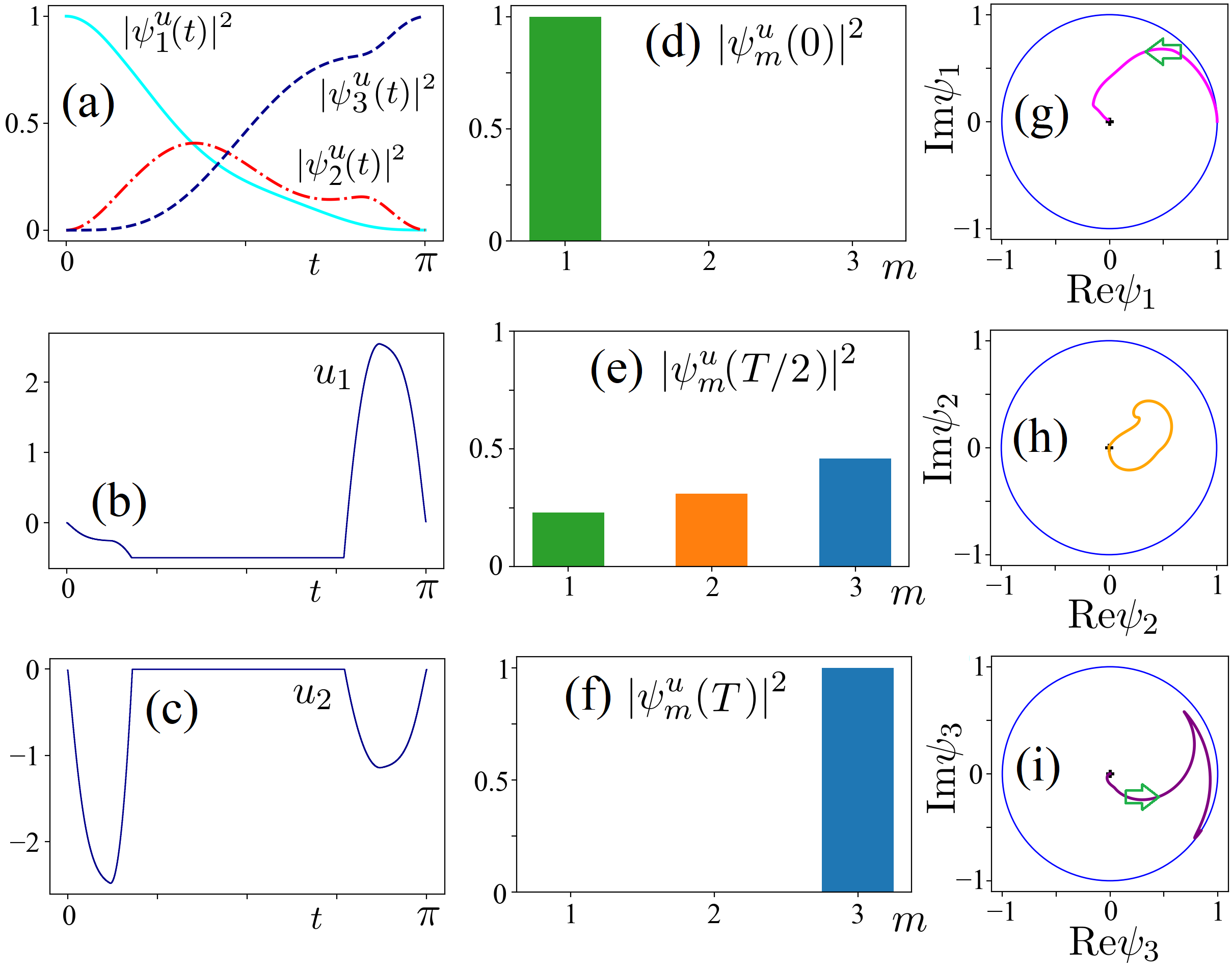}
\caption{For the transfer problems with the PConst. approximation of the special simple  class~(\ref{sect2_f6}), the GA results (Example~3) in the terms of the resulting $u_1, u_2$ (subfig.~(b,c)), and $\psi^u_j$, $j=1,2,3$ (subfig. (a), (d)--(i)).}
\label{Fig3} 
\end{figure} 

\vspace{0.15cm}

{\bf Example 3.}
Consider the transfer problem with (\ref{sect2_f10}). Take $T=\pi$, $P_{\bf x}=0$. In (\ref{sect2_f4}), take $b_1, b_2$ with $\overline{b}_1 =5$, $\overline{b}_2 = 3$, $q_1=q_2=8$. Let $\theta_l^{\rm L}, \theta_l^{\rm R} \in [-1,1]$, $y_l \in [-0.1\overline{b}_l, 0.1\overline{b}_l]$, $l=1,2$, $\widehat{t}_1 \in [0.07 T, 0.13 T]$, $\widehat{t}_2 \in [0.17 T, 0.23T]$, $\widehat{t}_3 \in [0.77 T, 0.83 T]$, $\widehat{t}_4 \in [0.87 T, 0.93 T]$ for the class~(\ref{sect2_f6}) approximated at the uniform $\Theta_M$ with $M=1500$. For minimizing $f_3({\bf x})$, GA was used with an~automatically generated initial point~${\bf x}^{(0)}$ for each trial GA run. The corresponding Python program was written by the author using the known GA implementation~\cite{Solgi_Genetic_algorithm_Python} and various tools from {\tt NumPy}, {\tt SciPy}, {\tt Matplotlib}, etc., including, e.g., {\tt scipy.linalg.expm} for a~{\it Pad\'{e} approximation} with a~variable~order. In a~GA run, the shown in Fig.~\ref{Fig3} resulting control process was obtained giving $I_1 \approx 8 \cdot 10^{-4}$ (here, e.g., $I_1 \approx 0.01$ was reached after solving (via Lemma~2) 2009 Cauchy problems~(\ref{sect2_f1})). Repeating the GA approach, one can see various locations of $\psi^u_3(T)$ approximately at~$S^1(1)$.

\vspace{-0.1cm}
\section{Conclusion-I}
\label{Sect5}
\vspace{-0.4cm} 

~\par This article (I) adds the certain pointwise constraints on controls, etc. to the transfer problem, introduces the keeping problem. For these problems under PContin. controls, the projection-type linearized PMP, GPM constructions are adapted (Lemma~1, Theorem~1, etc.). Lemma~2 adapts the known matrix exponentials' approach. With (\ref{sect2_f6}) and Lemma~2, GA is successfully used in Example~3 for the transfer problem with~$N=3$.

\vspace{0.15cm}

{\bf Acknowledgments.} The author is grateful: (a)~to A.S.~Buldaev (Ulan-Ude), Dr. Phys.-Math. Sci., Prof., who in 2005 and later  interested the author in the direction of projection-type methods in MTOC; (b)~to A.I.~Tyatyushkin (Irkutsk), Dr.~Tech. Sci., Prof., the author of \cite{Tyatyushkin_book_1992}, who in 2005 and later interested the author in the direction of 
fin.-dim. optimization in MTOC; (c)~to V.F.~Krotov (Moscow), a~Honored Scientist of the Russian Federation, Dr.~Tech. Sci., Prof., the author of \cite{Krotov_book_1996}, who in 2010 interested the author to consider~\cite{Caneva_Murphy_Calarco_et_al_2009}; (d)~to A.N.~Pechen (Moscow), Dr. Phys.-Math. Sci., RAS Prof., who several years ago drew the author’s attention to~\cite{Khaneja_et_al_2005, Pechen_Tannor_2012}\,(GRAPE).

\end{document}